\documentclass[a4paper, fleqn, oneside,10pt]{article}

\usepackage{amsmath}
\usepackage{amsfonts}
\usepackage{amsthm}

\DeclareMathOperator{\diverg}{div}

\DeclareMathOperator{\nlog}{ln}

\DeclareMathOperator{\tr}{Tr}
\DeclareMathOperator{\I}{I}

\newtheorem{lemma}{Lemma}[section]
\newtheorem{theorem}{Theorem}[section]
\newtheorem{proposition}{Proposition}[section]

\newtheorem{corollary}{Corollary}[section]
\newtheorem{definition}{Definition}[section]

{\catcode `\@=11 \global\let\AddToReset=\@addtoreset}
\AddToReset{equation}{section}

\newcommand{\ie}{{\sl i.e.\/ }}
\newcommand{\cf}{{\sl c.f.\/ }}
\newcommand{\eg}{{\sl e.g.\/}}

\newcommand{\newpar}{\par}\parindent =0pt\parskip=3pt\textheight = 615pt
\renewcommand{\L}{{L}}
\newcommand{\Id}[1]{{\rm I\kern-2pt I_{#1}}}
\renewcommand{\hbar}{{\displaystyle\bar{\phantom{x}}\kern-6pt h}}
\begin{document}
\thispagestyle{empty}
\begin{center}
\fontsize{20}{22}
\selectfont
\textbf{On the Long Time Behavior of the Quantum Fokker-Planck equation}
\end{center}
\vskip 1 cm
\begin{center}
\fontsize{12}{12}
\selectfont
C. Sparber\footnote{Inst. f. Mathematik, Univ. Wien, Botzmanngasse 9, A-1090 Vienna, Austria,\\
e-mails: christof.sparber@univie.ac.at, peter.markowich@univie.ac.at,},
J. A. Carrillo\footnote{Departamento de Matem\'atica Aplicada, Univ. de Granada, 18071-Granada, Spain,\\
e-mail: carrillo@ugr.es,}, 
J. Dolbeault\footnote{Ceremade, Univ. Paris IX-Dauphine, Place de Lattre de Tassigny 75775, Paris, France,\\
e-mail: dolbeault@ceremade.dauphine.fr}, 
P. A. Markowich$^1$
\end{center} 
\vskip 1 cm
\begin{center}
\textbf{Abstract}
\end{center}
\emph{We analyze the long time behavior of transport equations for a class of dissipative quantum systems with Fokker-planck type scattering operator, 
subject to confining potentials of harmonic oscillator type. We establish the conditions under which there exists a thermal equilibrium state and prove 
exponential decay towards it, using (classical) entropy-methods. Additionally, we give precise dispersion estimates in the cases were no equilibrium state exists.}

\section{Introduction}

In this paper we analyze a class of \emph{dissipative quantum systems}, modeling the motion of a particle ensemble, say electrons, 
interacting with a heat bath of oscillators. The resulting irreversible dynamics for the electrons, is a typical example of a so-called 
\emph{open quantum system} \cite{Da}, \ie a system in which the interaction with the environment is taken into account. 
The evolution equation, sometimes called \emph{Master equation}, for the
density (matrix) operator $R(t)$ of the particles reads
\begin{align}
\label{maeq}&\frac{d}{dt} R = -\frac{i}{\hbar } \left[H, R\right] + A(R),\\
&R(t=0)=R_0,
\end{align}
where $H$ is the self-adjoint \emph{Hamiltonian operator} of the
free system and $A$ models the effects introduced by the heat bath. The
right hand side of (\ref{maeq}) constitutes the formal generator $L$ of
the \emph{quantum dynamical semigroup} acting on $R$.\newpar Assuming that the
time evolution of the system is
\emph{Markovian}, G. Lindblad \cite{Li} gave the most general form of a
bounded operator $L$, such that the semigroup preserves the positivity,
hermiticity and the normalization (unit trace) of the density operator
$R$. However, for unbounded operators $L$, which is the case in our
work, the so-called \emph{Lindblad condition} is necessary but \emph{not
sufficient} to guarantee the conservation of these properties (see, \eg,
\cite{CF} and the references therein). \newpar Using the \emph{Wigner
transform} \cite{Wi}, dissipative quantum models can be equivalently
represented in \emph{phase space}, resulting in a \emph{kinetic
transport equation} with interaction terms for the \emph{quasiprobability
distribution} of the particles. In this paper we assume that the
mechanism coupling particles and environment can be described by a linear
scattering operator $\mathcal L_q$ of \emph{Fokker-Planck type}. The
Lindblad condition therefore reduces to the assumption of a
\emph{positive definite diffusion matrix} $D$.\newpar Up to now, a
mathematically rigorous derivation of such a \emph{Quantum Fokker-Planck
equation} (QFP) from many-body quantum mechanics is still missing. To
the authors knowledge, the only result in this direction is given in
\cite{CEFM}, which however justifies only a particular case of the class
of models considered in this work. Nevertheless there exists a huge
amount of a somewhat phenomenological physical literature on this type of
equations, which play a relevant role within the areas of quantum optics
(laser physics) \cite{Da}, \cite{De}, \cite{Ri}, microelectronics
\cite{St}, quantum brownian motion \cite{CaLe}, \cite{Di}, \cite{HuMa},
and the description of decoherence and diffusion of quantum states
\cite{AnHa}, \cite{DGHP}.\newpar Rigorous well-posedness and existence of
local in time solutions of the frictionless QFP equation, with
self-consistent Coulomb interaction, have been studied in a precedent
paper of one of the authors \cite{ALMS}. The present work investigates
the long time behavior of the linear QFP equation in the presence of
friction and an exterior time-independent potential $V$. The phase space
formalism provided by the Wigner transform proves to be particularly
helpful for this task, since it allows the use of certain
\emph{entropy techniques} established for classical dissipative
equations (for an overview on these techniques, see, \eg, \cite{MaVi},
\cite{AMTU}). The word "entropy" is used here in a mathematically
technical sense and can be seen as a generalization of the classical
entropy concept of L. Boltzmann. It should \emph{not} be confused with
the physical correct \emph{von Neumann entropy} of quantum states.\newpar
By comparison with the classical \emph{Fokker-Planck equation} (FP), we
expect the solution of the QFP equation to approach a thermal
\emph{equilibrium state} in the long time limit, provided the friction
term appearing in $\mathcal L_q$ is positive and the exterior potential
$V$ is \emph{confining}, 
\ie $V(x)\rightarrow \infty$ as $|x|\rightarrow \infty $, fast enough. 
In our work we shall specify the potential $V$ to be \emph{harmonic}. This particular choice allows explicit calculations and is maybe the most fundamental one, 
from a physical point of view \cite{CEFM}. 
Using the entropy approach, we will prove the convergence of the solution
towards the steady state with a precise exponential rate, under the
assumption that the initial data $w_0$ has bounded "entropy," relatively
to the equilibrium state. \newpar
This paper is organized as follows. In
section 2 we set up the model and collect some preliminaries.  In section
3 we specify the potential $V$ to be of harmonic oscillator type and
explicitly calculate the corresponding equilibrium state. Exponential
decay towards it will be proved in section 4, where we also give precise
dispersion estimates in the unconfined cases. 
 

\section{The model: preliminaries}
\noindent We consider a linear dissipative equation modeling the motion of
particles, say electrons, under the influence of  an electric scalar
potential
$V$ and a thermal bath of harmonic oscillators in thermal equilibrium. 
In the following we denote by $\rho (t,\cdot )\in L^2(\mathbb R^{2d})$
the N-particle \emph{density matrix function} of the electrons,  which
is the kernel of the self-adjoint trace-class 
\emph{density (matrix) operator} $R(t): L^2(\mathbb R^d)\rightarrow L^2(\mathbb R^d)$,
\ie
\begin{equation}
\label{dop} (R(t)f)(x):=\int_{\mathbb R^d} \rho (x,y,t)f(y)dy.
\end{equation} 
The evolution equation of the electrons is given as a PDE for the density matrix 
\begin{align}
\label{evol}\partial_t\rho =&-\frac{i}{\hbar }(H_x-H_y)\rho -\gamma (x-y)\cdot (\nabla _x-\nabla _y)\ \rho \nonumber \\
&+\left(D_{qq} |\nabla _x+\nabla _y|^2-\frac{D_{pp}}{\hbar ^2}|x-y|^2+\frac{2iD_{pq}}{\hbar }(x-y)\cdot (\nabla _x+\nabla _y)\right)\rho ,
\end{align}
subject to the initial condition
\begin{align}
\rho (t=0,x,y)=\rho _0(x,y),\quad x,y\in\mathbb R^d.
\end{align}
Here $H_{x}$ (resp. $H_{y}$) denotes the electron \emph{Hamiltonian} 
\begin{equation}
H_x:=-\frac{\hbar ^2}{2m}\Delta_x +V(x),
\end{equation}
acting on the $x$ (resp. $y$) variable. The constant $m$ stands for the
mass of the individual particles. Equation (\ref{evol}) is a
generalization of the Caldeira-Leggett master equation for medium
temperatures, see \cite{Di},
\cite{De}.

\noindent On a kinetic level this model reads (QFP equation)
\begin{align}
\label{qfp}\partial_tw +\xi \cdot\nabla _xw +\Theta [V] w = & \ \mathcal L_{q}w \quad\quad x,\xi \in\mathbb R^d, t\in\mathbb R^+ \\ 
\label{ico}w (t=0,x,\xi ) = & \ w _0(x,\xi ),
\end{align}
where the scattering operator $\mathcal L_q$ is defined by
\begin{equation}
\label{lq}\mathcal L_{q}w :=\ \frac{D_{pp}}{m^2} \Delta _\xi w +2\gamma \diverg_\xi (\xi w ) + 
D_{qq} \Delta _xw +2\frac{D_{pq}}{m} \diverg_x(\nabla _\xi w ).
\end{equation}
Here $w (t,\cdot ) \in L^2 (\mathbb R^d_x\times \mathbb R^d_\xi )$ is the
\emph{Wigner transform} \cite{Wi} of the corresponding  density matrix
$\rho(t,\cdot ) \in L^2(\mathbb R_x^d\times \mathbb R^d_y )$, \ie 
\begin{equation}
w (x,\xi ,t):=\frac{1}{(2\pi )^d\hbar}\int_{\mathbb R^d}\rho \left( x+\frac{\hbar }{2m}\ y, \ x-\frac{\hbar }{2m}\ y, t\right)e^{i\xi \cdot y}dy.
\end{equation}
In the literature $w(t,\cdot ) $ is often referred to as a quasiprobability 
distribution because it generally assumes negative values too \cite{Fo}, \cite{Hu}. \newpar 
The (non-local) pseudo-differential operator $\Theta [V]$ is
defined by
\begin{align}
\Theta [V]f(x,\xi) := \frac{i}{(2\pi )^d \hbar }\int\int_{\mathbb R^d\times\mathbb R^d}
\left[V\left({ x+\frac{\hbar }{2m}\ y}\right)-V\left({ x-\frac{\hbar
}{2m}\ y}\right)\right] \nonumber\\  f(x,\xi')\ e^{iy\cdot (\xi
-\xi')}d\xi 'dy .
\end{align}

\noindent In order to be consistent with the usual density matrix formulation of \emph{open quantum systems} \cite{Da} in the class of \emph{Lindblad operators} \cite{Li}, we assume for the \emph{diffusion constants} $D_{pp}>0$ and $D_{qq}, D_{pq}\geq
0$. On the other hand, the \emph{friction parameter} $\gamma$ has to be
nonnegative. Additionally we impose the following relation (for more
details see
\cite{ALMS}, \cite{Li})
\begin{equation}
\label{li}D_{pp}D_{qq}-D_{pq}^2 \geq\frac{\hbar ^2\gamma ^2}{4}\quad\mbox{and}\quad D_{pp}>0\;\mbox{if}\; \gamma=0.\tag{\textbf{A1}}
\end{equation}
Note that if $\gamma >0$, condition (\ref{li}), the so-called
\emph{Lindblad condition}, implies that (\ref{lq}) is a
uniformly elliptic operator. 

\noindent Using the Wigner transform the \emph{charge} and \emph{flux densities} associated to the density matrix $\rho $ can be defined 
(formally, since $w\not\in L^1$ in general) in the same way as in classical statistical
mechanics. Namely they are given as moments of the Wigner transfom,  (for
details see, \eg, \cite{GaMa}, \cite{LiPa}),
\begin{align}
&n(x,t):=\int_{\mathbb R^d}w(x,\xi ,t)d\xi ,\\
\label{flux}&j(x,t):=\int_{\mathbb R^d}\xi w(x,\xi ,t)d\xi .
\end{align}
Although not obvious from the above definition, the necessary positivity of $n$ is guaranteed for \emph{physical quasiprobabilities} $w$, \ie for 
$w$'s, which are indeed the Wigner transformed kernels $\rho$ of density operators $R$ 
(see \cite{Ar}, \cite{LiPa}, \cite{Ta} for a more complete account on this).\newpar 
With the above definitions we obtain, after formally integrating the QFP
equation (\ref{qfp}) w.r.t. $\xi $, the associated "\emph{continuity
equation}"
\begin{equation}
\label{cont}\partial _tn+\diverg j=D_{qq}\Delta _xn,
\end{equation}
which obviously implies (again on a formal level) the \emph{conservation of mass}, \ie
\begin{equation}
M(t):= \int\int_{\mathbb R^{d}\times \mathbb R^d}w (x,\xi ,t)dxd\xi =\int\int_{\mathbb R^{d}\times \mathbb R^d}w_0 (x,\xi )dxd\xi.
\end{equation}
In view of this property, we assume for simplicity that
\begin{equation}
M_0= \label{nc}\int\int_{\mathbb R^{d}\times \mathbb R^d}w_0 (x,\xi )dxd\xi=1.\tag{\textbf{A2}}
\end{equation}
This is of course not a restriction as long as the equation is linear.
The continuity equation (\ref{cont}) suggests that in this dissipative model the usual definition of the flux density (\ref{flux}) needs to be replaced by 
\begin{align}
J(x,t):&=j(x,t)-D_{qq}\nabla _x n(x,t)\\
&=\int_{\mathbb R^d}(\xi -D_{qq} \nabla _x) \ w(x,\xi ,t)d\xi. 
\end{align}
Thus, instead of (\ref{cont}), we obtain the following conservation law
\begin{equation}
\label{cont2}\partial _tn+\diverg J=0
\end{equation}
associated to the QFP equation.

\noindent \textbf{Remark.} In physical units, the friction and diffusion
constants are usually given by (see, \eg, \cite{Di}, \cite{De})
\begin{equation}
\label{pun}\gamma =\frac{\lambda }{2m}, \ D_{pp} = \lambda k_BT, \ D_{qq} = \frac{\lambda \hbar ^2}{12m^2k_BT}, \ 
D_{pq}=\frac{\lambda \Omega \hbar ^2}{12\pi mk_BT}.
\end{equation}
Here $\lambda >0$ is the coupling constant of the heat bath, $k_B$ the Boltzmann constant, $T$ the temperature of the bath and $\Omega $ the cut-off frequency 
of the reservoir oscillators. In terms of these constants the Lindblad condition (\ref{li}) reads
\begin{equation}
\Omega \leq \frac{k_BT}{\hbar }
\end{equation}
which implies the validity of our model at medium/high temperatures.

\noindent In the classical limit $\hbar \rightarrow 0$: $D_{qq} = D_{pq}= 0$ and (at least formally) the pseudo-differential operator simplifies to 
\begin{equation}
\Theta [V] f\rightarrow -\frac{1}{m}\nabla _xV\cdot\nabla _\xi f. 
\end{equation}
Thus we recover the well known kinetic Fokker-Planck equation for 
the limiting classical phase space probability distribution
$w^{cl}(t,\cdot )\in\mathcal M^+(\mathbb R^d_x\times \mathbb R^d_\xi )$ 
(the cone of positive bounded Borel measures)
\begin{equation}
\label{fp}\partial_tw^{cl} +\xi \cdot\nabla _xw^{cl} - \frac{1}{m}\nabla _xV\cdot\nabla _\xi w^{cl} = \ \frac{D_{pp}}{m^2} \Delta _\xi w^{cl} +2\gamma \diverg_\xi (\xi w^{cl}). 
\end{equation}
Also, the flux density $J$ formally simplifies to the classical one, \ie $J\rightarrow j$, as $\hbar \rightarrow
0$. For a rigorous theory of such \emph{homogenization limits} see, \eg,
\cite{LiPa}, \cite{GMMP} and  for details on the extensively studied FP
equation, see for example~\cite{Ri}.\newpar 
Furthermore note that the Lindblad condition (\ref{li}) disqualifies the
classical FP scattering operator \cite{CEFM} as a relevant quantum
mechanical model for the environment interaction.

\noindent In this work we are concerned with the following solution concept for our IVP.
\begin{definition} A function $w \in C(\mathbb R_0^+; L^p( \mathbb R^d_x\times \mathbb R^d_\xi ) )$, with $1\leq p<\infty $, is a {\rm mild
solution} of the IVP (\ref{qfp}), (\ref{ico}) with $w_0\in\L^p(\mathbb R^d_x\times \mathbb R^d_\xi )$, if and only if
\begin{equation}
\label{ws}w (t,x,\xi )=\int\int_{\mathbb R^d\times \mathbb R^d}w _0(x_0,\xi _0) \ G (t,x,\xi ,x_0,\xi _0) \ dx_0d\xi _0,
\end{equation}
where the Green's function $G$ satisfies equation (\ref{qfp}) for all fixed $(x_0,\xi _0)\in\mathbb R^{2d}$, all
$t>0$, with an initial condition
\begin{align}
\lim_{t\searrow 0} G(t,x,\xi,x_0, \xi_0 )=\delta (x-x_0,\xi-\xi _0 ).
\end{align}
that has to be understood in a weak sense.
If additionally $w\in C^1(\mathbb R^+;C^2(\mathbb R^d_x\times \mathbb
R^d_\xi ) )$, then $w$ is called a {\rm classical solution.}
\end{definition}
\vskip 0.3 cm
\noindent \textbf{Remark.} In classical physics the theory of kinetic equations focuses on $L^1$-solutions. 
This implies mass conservation, since classical phase-space distributions are pointwise positive functions. 
Having in mind the definition of the Wigner transform, the $L^2$-norm is more convenient in our quantum mechanical context. 


\section{Harmonic oscillator potentials}

\noindent In the next two sections we choose a normalization such that $\hbar =m=1$, for
simplicity. We moreover assume the confining potential to be of the
following class
\begin{equation}
\label {hop}V(x)= \frac{\omega^2 _0}{2}|x|^2+ax+b,\quad a,b\in\mathbb R,\ \omega _0\geq 0.\tag{\textbf{A3}}
\end{equation}
An easy calculation shows that, maybe after an appropriate shift in the $x$-variable, the pseudo-differential operator $\Theta [V]$ is given by
\begin{equation}
\Theta [V]w =-\omega^2 _0x\cdot\nabla _\xi w.
\end{equation}
The QFP equation thus simplifies to 
\begin{align}
\label{qfph}\partial_tw +\xi \cdot\nabla _xw -\omega^2 _0x\cdot \nabla _\xi w = & \ \mathcal L_{q}w \quad\quad x,\xi \in\mathbb R^d, t\in\mathbb R^+\\ 
w (t=0,x,\xi ) = & \ w _0(x,\xi ),
\end{align}
which can be equivalently written in the more compact form
\begin{align}
\label{qfphre}&\partial _t w = \ \diverg_{(x,\xi) } \left(D \nabla_{(x,\xi) } w+P(x,\xi )w\right) , \\ 
&w (0,x,\xi )= \ w _0(x,\xi ),
\end{align}
where the \emph{diffusion matrix} $D$ and the vector-valued \emph{drift} $P$ are given by
\begin{equation}
\label {dp}D:=
\begin{pmatrix}
D_{qq}\Id{d} & D_{pq}\Id{d}\\
D_{pq}\Id{d} & D_{pp}\Id{d}
\end{pmatrix}
,\quad P(x,\xi ):=
\begin{pmatrix}
-\xi \\
\omega^2 _0x+2\gamma \xi 
\end{pmatrix}.
\end{equation}
Here $\Id{d}$ denotes the idendity matrix in $\mathbb R^d$. 
Note that the Lindblad condition (\ref{li}) guarantees that the diffusion matrix is positive definite if $\gamma >0$ and 
then (\ref{qfph}) is parabolic in the phase-space coordinates $(x,\xi )\in\mathbb
R^{2d}$.

\subsection{Fundamental solution}

\noindent Now consider only the first order part of the operator (\ref{qfph}), resp. (\ref{qfphre}). The associated characteristic ODE's are given by
\begin{align}
\label{code1} &\dot X=\xi , &X(t=0)=x_0,\\
\label{code2} &\dot \xi =-(\omega^2 _0X+2\gamma \xi ), &\xi (t=0)=\xi _0.
\end{align}
This system defines the characteristic flow $\Phi_t(x_0,\xi _0)=[X_t (x_0,\xi _0),\dot X_t (x_0,\xi _0)]$ in phase space $\mathbb R^d_x\times \mathbb R^d_\xi $. This flow can be explicitly calculated, depending on the size of the friction constant $\gamma $.

\begin{lemma} Consider the system (\ref{code1}), (\ref{code2}) with $\gamma \geq 0$, then the dissipative flow 
$\Phi _t:\mathbb R^{2d}\rightarrow \mathbb R^{2d}$ is given by: 
\begin{enumerate}
\item If $\ 0\leq \gamma <\omega _0$, then, with $\omega :=\sqrt{\omega^2
_0-\gamma^2 }$, 
\begin{align}
\label{flow1}\Phi _{t}(x_0,\xi_0 )= \frac{e^{-\gamma t}}{\omega }\big[& x_0 (\omega \cos (\omega t)+\gamma \sin(\omega t))+\xi_0 \sin(\omega
t),\nonumber\\
& \xi _0 (\omega \cos(\omega t)-\gamma \sin(\omega t)) -x_0\omega _0^2 \sin(\omega t)\big].
\end{align}
\item If $\gamma>\omega _0$, then, with $\omega :=\sqrt{\gamma^2-\omega^2
_0 }$, 
\begin{align}
\label{flow2}\Phi _{t}(x_0,\xi_0 )= \frac{e^{-\gamma t}}{\omega}\big[ & x_0 (\omega \cosh (\omega t)+\gamma \sinh (\omega t))+
\xi _0 \sinh(\omega t),\nonumber\\
& \xi _0 (\omega \cosh (\omega t)-\gamma \sinh (\omega t)) -x_0 \omega _0^2 \sinh(\omega t) \big].
\end{align}
\item If $\gamma=\omega _0$ then 
\begin{equation}
\label{flow3}\Phi _{t}(x_0,\xi_0 )= e^{- \gamma t}\big[(\gamma t+1)x_0+t\xi _0, \ (1-\gamma t)\xi _0-\gamma ^2 t x_0\big].
\end{equation}
\end{enumerate}
\end{lemma}
\begin{proof}
The proof follows from straightforward calculations. 
\end{proof}
\noindent Now, using lemma 2.1, we obtain an explicit representation of Green's
function~$G $ of the QFP equation with harmonic oscillator potential,
depending on the size of $\gamma $. 
\begin{proposition}
Let $\gamma \geq 0$ and let conditions (\ref{li}), (\ref{hop}) hold, then the Green's function $G
$ associated to (\ref{qfph}) is,  for every fixed $t>0$, a pointwise
positive function, given by
\begin{equation}
\label{vartr}G (t,x,\xi ,x_0,\xi _0) = e^{2d\gamma t}F (t,X_{-t}(x,\xi )-x_0,\dot X_{-t}(x,\xi )-\xi _0),
\end{equation}
with
\begin{equation}
F (t,x,\xi )= \frac{\exp \left(- \frac{ \nu(t)\vert x\vert^2+\lambda(t)\vert
\xi \vert^2+\mu(t)(x\cdot \xi)}{4\lambda(t)\nu(t)-\mu^2(t)} \right)}{( 2\pi )^{d}
(4\lambda(t)\nu(t)-\mu^2(t))^{d/2}}\in C^1(\mathbb R^+;\mathcal S(\mathbb R^d_x\times \mathbb R^d_\xi )).
\end{equation}
In (\ref{vartr}) we denote by $X_{-t}, \dot X_{-t}$ the components of the inverse characteristic flow $\Phi _{-t}$, which satisfies $\Phi
_{-t}\circ\Phi _{t}= {\rm id}$:
\begin{equation*}
\Phi_{-t}(x,\xi) = [X_{-t} (x,\xi),\dot X_{-t} (x,\xi)] .
\end{equation*}
The associated functions $\lambda, \nu , \mu:\mathbb R^+\rightarrow \mathbb R $ are defined by the following expressions:
\begin{align}
\lambda (t)&:=\int_0^t \Bigl( D_{qq}\alpha^2(s)+D_{pp}\beta^2(s)+2D_{pq}\alpha(s)\beta(s)\Bigr)ds,\\
\nu (t)&:=\int_0^t \left( D_{qq}\dot\alpha^2(s)+D_{pp}\dot\beta^2(s)+2D_{pq}\dot\alpha(s)\dot\beta(s)\right)ds,\\
\mu (t)&:=2\int_0^t \left( D_{qq}\alpha(s)\dot\alpha(s)+D_{pp}\beta(s)\dot\beta(s)+D_{pq}\frac d{ds}{(\alpha(s) \beta(s))}\right)ds,
\end{align}
where the functions $\alpha ,\beta:\mathbb R^+\rightarrow \mathbb R $ are given by:
\begin{align}
\alpha (t) := d^{-1}\diverg_x \left(X_{-t}(x,\xi )\right),\quad \beta (t) := d^{-1}\diverg_\xi \left(X_{-t}(x,\xi )\right).
\end{align}
\end{proposition}
\vskip 0.3 cm
\begin{proof} The proof is similar to the calculations given in \cite{Bo}, \cite{Ho}. First note that by definition of $\alpha$ and $\beta$,
we can write $X_{-t}(x,\xi )=\alpha (t)x+\beta (t)\xi$. Thus if $G $ is the fundamental solution of (\ref{qfph}), the linear transformation
(\ref{vartr}) guarantees that $F $ has to be a fundamental solution of the following PDE with time dependent coefficients
\begin{equation*}
\partial_t
F (t,x,\xi )=\left(\frac{d\lambda}{dt}(t) \Delta _x+\frac{d\nu}{dt}(t)\Delta _\xi -\frac{d\mu}{dt}(t)\nabla _x\cdot \nabla _\xi \right)F (t,x,\xi ) .
\end{equation*}
where the functions $ \lambda , \nu , \mu$ are calculated depending on the choice of $\gamma $. A Fourier transform now shows that
\begin{equation*}
(\mathcal F F)(t,k,\eta )\equiv \hat F (t, k,\eta):=\int\int_{\mathbb R^d\times \mathbb R^d}F (t,x,\xi )
e^{-i(x \cdot k+ \xi\cdot\eta )}\; dxd\xi 
\end{equation*}
is a solution of
\begin{equation*}
\partial_t
\hat F (t,k,\eta )=-\left(\frac{d\lambda}{dt}(t) |k|^2 +\frac{d\nu}{dt}(t) |\eta |^2 -\frac{d\mu}{dt}(t) (k\cdot\eta )\right) \hat F (t,k,\eta ).
\end{equation*}
This equation can easily be integrated and thus, using that $\hat F (t,0,0)=1$ for all $t\geq 0$, we obtain
\begin{align*}
\nlog \left(\hat F (t,k,\eta)\right)&=
-\left(\lambda(t)\vert k\vert^2+\nu(t)\vert \eta \vert^2 -\mu(t) (k\cdot \eta ) \right).
\end{align*}
After some lengthy and tedious calculations (where one checks that 
$4\lambda \nu \geq \mu ^2$), an inverse Fourier transform gives 
\begin{align*}
F (t,x,\xi )& =(2\pi )^{-2d} \int\int_{\mathbb R^d\times \mathbb R^d}
e^{i(x \cdot k+\xi \cdot \eta )}\hat{F }(t,k,\eta )dkd\eta \\
&=(2\pi )^{-d} \frac{e^{-\frac{\vert
x\vert^2}{4\lambda}}}{(4\pi \lambda)^{d/2}}
\int_{\mathbb R^d}e^{-i \frac{\mu}{2\lambda}
(x\cdot\eta)-\frac{1}{4\lambda}(4\lambda\nu-\mu^2)\vert
\eta\vert^2+i \xi \cdot \eta} d\eta\\
&=\frac{\exp{\left(-\frac{\nu(t)\vert x\vert^2+\lambda(t)\vert
\xi \vert^2 + \mu(t)(x\cdot \xi) }{4\lambda(t)\nu(t)-\mu^2(t)}\right)}} {(2\pi )^d
(4\lambda(t)\nu(t)-\mu^2(t))^{d/2}},
\end{align*}
which is the desired result.
\end{proof}
\noindent Note that the quantum mechanical effects in $F $ and consequently in $G $ only enter in form of the constants $D_{qq}, D_{pq}\sim \hbar ^2$, which appear in the auxiliary functions $\lambda ,\nu ,\mu $. In other words we
obtain Green's function for the classical Fokker-Planck equation in a square-well potential by setting $D_{qq} =D_{pq} =0$ in the above
expressions. The pointwise positivity of $G$ is a consequence of the minimum principle for parabolic equations of Fokker-Planck type
\cite{Ev}.

\noindent From the above proposition we draw the following consequences (among which we obtain the conservativity of the quantum dynamical
semigroup, associated to the harmonically confined QFP equation).

\begin{corollary} Let $\gamma \geq 0$ and assume (\ref{li})-(\ref{hop}). Then for every initial condition $w _0\in L^p(\mathbb R^d_x\times \mathbb R^d_\xi )$ 
with $1\leq p<\infty$, there exists a unique classical solution $w \in C(\mathbb R_0^+; L^p(\mathbb R^d_x\times \mathbb R^d_\xi ))$ $\cap$ 
$C^1(\mathbb R^+; C^\infty (\mathbb R^d_x\times \mathbb R^d_\xi ))$ with $M(t)=M_0\equiv 1$. 
Moreover if $w _0$, $n_0$ are non-negative a.e., so are $w (t,\cdot )$, $n(t,\cdot )$, for all $t\in\mathbb R^+$.
\end{corollary}
\begin{proof} For convenience we use the notation: $y:=(x,\xi )$ as well as $y_0:=(x_0,\xi _0)$. With the following linear change of variables 
\begin{equation*}
G (t,\Phi _t(y),y_0)= e^{2d\gamma t}F (t,y-y_0)\in C^1(\mathbb R^+, \mathcal S(\mathbb R^{2d})),
\end{equation*}
we can express our solution in the form
\begin{equation*}
w (t, \Phi _t(y))=e^{2d\gamma t}(w_0 \ast F (t,y)).
\end{equation*}
A straightforward computation now shows that the Jacobian determinant of the mapping $\Phi _t(\cdot )$ is given by
\begin{equation*}
\label{jac}\det \left(\frac{\partial \Phi _t(y)}{\partial y}\right)=\exp(-2d\gamma t).
\end{equation*}
With these preparations and using Young's inequality \cite{LiLo}, we obtain
\begin{align*}
{\| \ w (t)\ \|} _p={\| \ w_0 \ast F (t)\ \| }_p
\leq {\| \ w _0 \ \|} _p \ {\| \ F (t)\ \|} _1<\infty ,
\end{align*}
since for each fixed $t\in\mathbb R^+$: $F (t,\cdot )\in\mathcal S(\mathbb R^{2d})\subset L^1(\mathbb R^{2d})$. More precisely we have
\begin{equation*}
{\| F (t)\parallel }^p_p=\int_{\mathbb R^{2d}}|F (t,y)|^p \ dy= p^{-1},
\end{equation*}
for all $1\leq p<\infty$. For $p=1$ this implies, after a simple calculation, that $M(t)=M_0\equiv 1$, by (\ref{nc}). 
Since $G (t,\cdot )$ is pointwise positive, we clearly obtain that if $w_0 $, resp. $n_0$ are a.e. non-negative, so are $w (t,\cdot )$,
$n(t,\cdot )$. 
\end{proof}

\noindent\textbf{Remark.} If the initial condition $w _0$ is the Wigner transform of a pure quantum state $\psi_0$, it is well known \cite{Hu}, 
\cite{LiPa} that $w_0 \geq 0$ pointwise, if and only if $\psi_0 $ is a Gaussian. A similar characterization for mixed states has not been
found~yet.

\noindent From the above result it is easy to deduce the
\emph{conservativity} (\ie conservation of hermiticity,  positivity and
normalization of the density matrix $\rho$) of the quantum dynamical
semigroup corresponding to (\ref{qfph}), using  the inverse Wigner
transform 
\begin{equation}
\rho (x,y,t)= (2\pi )^{-d} \int_{\mathbb R^d} w\left(\frac{x+y}{2}, \xi,t \right)e^{i\xi \cdot (x-y)} d\xi,
\end{equation}
which is defined in the sense of the usual $L^2$-Fourier transform
\cite{Fo}, \cite{LiLo}.

\subsection{Stationary states.}

\noindent As in the classical case, we expect that the competing effects of the confining potential plus the positive friction 
and the dissipating behavior of the operator result in a \emph{thermal equilibrium state} 
as $t\rightarrow \infty $.
\begin{definition} A (thermal) equilibrium state is a steady state, \ie stationary solution $w\in L^2 (\mathbb R^d_x \times \mathbb R^d_\xi )$ 
of the QFP equation (\ref{qfp}), such that 
\begin{equation}
J(x)=\int_{\mathbb R^d}(\xi -D_{qq} \nabla _x) \ w(x,\xi)d\xi=0. 
\end{equation}
\end{definition}
\noindent In principle it could be possible that there exist stationary solutions of the QFP equation, which are not equilibrium states, 
however the following proposition and corollary show that this is not the case.\newpar 
Furthermore, we shall see that the physical intuitive assumption, that the
mass of the initial state is equal to the mass of the steady state, 
guarantees its uniqueness. 
\begin{proposition} Let $\gamma > 0$, $\omega _0>0$ and assume (\ref{li}), (\ref{hop}), then the unique solution 
$w_\infty $ of the stationary QFP equation, satisfying 
\begin{equation}
M_\infty :=\int_{\mathbb R^{d}}\int_{\mathbb R^{d}} w_\infty (x,\xi ) \ dxd\xi =1,
\end{equation}
is given by the following non-isotropic Gaussian function
\begin{equation}
\label{qstst} w _\infty (x,\xi )=\frac{\gamma \omega _0}{(2\pi)^{d} \sqrt{Q}}\ 
\exp\left(-\frac{\gamma }{Q}\Bigl[Q_{11} \omega _0^2|x|^2+2Q_{12 } \omega _0x\cdot \xi +Q_{22 } |\xi |^2 \Bigr]\right),
\end{equation}
such that $w_\infty\in\mathcal S(\mathbb R^d_x \times \mathbb R^d_\xi )$, with $Q:=Q_{11}Q_{22}-Q^2_{12}$ and 
\begin{align}
Q_{11}&:=D_{pp}+\omega _0^2D_{qq},\\
Q_{12}&:=2\omega _0\gamma D_{qq},\\
Q_{22}&:=D_{pp}+\omega _0^2D_{qq}+4\gamma (D_{pq}+\gamma D_{qq}).
\end{align}
\end{proposition} 
\begin{proof} The proof is much simpler in Fourier-space where the steady state $\hat w_\infty $ is explicitly given by
\begin{align*}
\hat w _{\infty }(k,\eta )= \exp& \left( -\frac{D_{pp}}{4\gamma \omega _0^2}\ (|k|^2+\omega _0^2|\eta |^2)-\right.
\frac{D_{pq}}{\omega _0^2}\ |k|^2-\\
& \ \left. D_{qq}\big[(\frac{\gamma }{\omega _0^2}+\frac{1}{4\gamma }) \ |k|^2+\frac{\omega _0^2}{4\gamma } |\eta |^2+k\cdot \eta \big] \right).
\end{align*}
First it is straightforward to check that this is indeed a solution of the Fourier transformed stationary QFP equation
\begin{equation*}
\label{fst}\omega _0^2 \eta \cdot\nabla _k \hat w_\infty +(2\gamma \eta -k)\cdot \nabla _\eta \hat w_\infty = 
-\left (D_{pp} |\eta |^2\kern -0.5pt +\kern -0.5pt D_{qq}|k|^2\kern -0.5pt +\kern -0.5pt 2D_{pq}k\cdot \eta \right)\hat w_\infty 
\end{equation*}
and then that it satisfies the Fourier transformed mass normalization
condition $\int\!\!\int w_\infty dxd\xi =1$, \ie 
\begin{equation}
\label{fnc} \hat w_\infty (0,0)=1.
\end{equation} 
Thus, it remains to prove the uniqueness of the steady state. To do so we consider the corresponding characteristic system 
\begin{align}
\label{fode1}&\dot k=\omega _0^2 \eta,& k(s=0)=k_0\\
\label{fode2}&\dot \eta =2\gamma \eta-k,& \eta (s=0)=\eta _0.
\end{align}
Having in mind that, by assumption, $\gamma > 0$ we check that the real parts of the eigenvalues $\lambda _{1,2}:=\gamma
\!\pm\! \sqrt{\gamma ^2\!-\!\omega _0^2}$ of this system are positive,
which, by~the standard theory of ODE's, implies that $(k,\eta )=(0,0)$ is
a source of the characteristic flow (a similar argument holds if 
$\gamma =\omega _0$).
Thus the condition (\ref{fnc}) is necessary and sufficient to guarantee
the uniqueness of the solution~$\hat w _\infty $.
\end{proof}
\noindent From the above proposition we draw the following consequences:
\begin{corollary} Under the same assumptions as above we have:
\begin{enumerate}
\item The state $w_\infty $ is also an equilibrium state, \ie $J_\infty(x)=0$ for all $x\in\mathbb R^d$.
\item If either $\omega _0=0$ (free motion case) or $\gamma =0$ (frictionless case) or both are equal to zero, no nontrivial $L^1$-steady state
exists. 
\end{enumerate}
\end{corollary} 
\begin{proof} It is a lengthy but straightforward calculation to show that the current-density $J_\infty$ associated to $w_\infty $ vanishes identically. \newpar 
In the limiting cases $\gamma =0$ or $\omega _0=0$, we consider the characteristic curves given by the ODE system (\ref{fode1}), (\ref{fode2}). 
Along these curves the steady state varies according to 
\begin{equation}
\label{intchar}\frac{d}{ds}\hat w _\infty(k_s,\eta _s) = -\vartheta _s(k_0,\eta _0) \ \hat w _\infty(k_s,\eta _s) ,
\end{equation}
where $\eta _s=\eta _s(k_0,\eta _0)$, $k _s=k_s(k_0,\eta _0)$ denote the (vector valued) characteristic curves starting
for $s=0$ at the point $(k_0,\eta _0)$. By $\vartheta _s(k_0,\eta _0)$, we
mean
\begin{align*}
\vartheta _s(k_0,\eta _0):=D_{pp} | \eta_s (k_0,\eta _0) |^2+D_{qq}|k_s(k_0,\eta _0)|^2+2D_{pq}(k_s\cdot \eta_s) (k_0,\eta _0). 
\end{align*}
We integrate equation (\ref{intchar}) and obtain the Fourier transformed steady state parametrized by $s\in\mathbb R$
\begin{align}
\label{intsol}\hat w _\infty (k_s(k_0,\eta _0), \eta_s (k_0,\eta _0))=\hat w _\infty (k_0,\eta _0)\exp\left(-\int_0^s \vartheta_\tau (k_0,\eta _0)\ d\tau \right).
\end{align}
In case the potential vanishes, \ie $\omega _0=0$ (and $\gamma \geq 0$), we have for all $s\in\mathbb R$ that $k_s(k_0,\eta _0)=k_0$ 
(free motion) and
\begin{equation*}
 \eta_\tau (k_0,\eta _0) =
\begin{cases}
\eta _0-k_0\tau & \text{if $\gamma =0$},\\
\eta _0 \exp(2\gamma \tau )-\frac{k_0}{2\gamma } &\text{if $\gamma >0$.}
\end{cases}
\end{equation*}
Now let $\omega _0=\gamma =0$ and set $k_0=0$ in equation (\ref{intsol}). We obtain
\begin{equation*}
\hat w _\infty (0,\eta _0)=\hat w _\infty (0,\eta _0)\exp\left(-D_{pp}|\eta _0|^2s\right),\quad \forall \ \eta _0\in\mathbb R^d,s\in\mathbb R,
\end{equation*}
which implies 
\begin{equation*}
\hat w _\infty (0,\eta _0)=0,\quad \forall \ \eta _0\in\mathbb R^d,
\end{equation*}
in contradiction to (\ref{fnc}). The same type of argument holds in the case $\omega _0=0$, $\gamma >0$, if we set $\eta _0=k_0/2\gamma $ in equation 
(\ref{intsol}).\newpar 
In the frictionless case, \ie $\gamma =0$, $\omega _0>0$, the characteristics are circles in the $(k,\eta )$ plane. Thus for $s=2\pi / \omega _0$ we have
$(k_s(k_0,\eta _0),\eta _s(k_0,\eta _0))=(k_0,\eta _0)$.
However if we integrate over one period, \ie setting $s=2\pi /\omega _0$ in (\ref{intsol}), we obtain for all $k_0, \eta _0\in\mathbb R^d$
\begin{equation*}
\hat w _\infty (k_0,\eta_0)=\hat w _\infty (k_0,\eta _0)\exp\left(-\frac{\pi }{\omega^3 _0}(D_{pp}+\omega _0^2D_{qq}) \ 
(|k_0|^2+\omega _0^2|\eta _0|^2) \right) 
\end{equation*}
which clearly implies, since $D_{pp},D_{qq}>0$, that $\hat w_\infty $ identically vanishes. 
\end{proof}
\noindent The above corollary in particular shows that, as expected, the presence of a positive friction together with a confining potential
are crucial to guarantee the existence of an admissible, \ie with finite
mass, equilibrium state.\newpar  Note that although the solution
$w$ of (\ref{qfp}) in general will not be nonnegative, the  steady state
$w _\infty $ is nevertheless a pointwise \emph{positive} function,
because of Assumption (\ref{nc}).  The associated density matrix $\rho
_\infty $ is a particular example of a
\emph{mixed} quantum state with  positive Wigner transform. It is
explicitely given~by
\begin{equation}
\rho _\infty(x,y)=\frac{\gamma \omega _0}{(16\pi^3 )^{d/2}\sqrt{\gamma Q_{22}}}\ 
e^{-\frac{1}{4\gamma Q_{22}}\left[\gamma ^2\omega _0^2(x+y)^2+Q(x-y)^2\right]+i \omega _0 \frac{Q_{12}}{ Q_{22}}\
\left(\frac{x^2-y^2}{2}\right)} .
\end{equation}
Observe that the equilibrium density $n_\infty $, which is obtained from the diagonal (\ie $x=y$) of the steady state density matrix $\rho _\infty $, 
is a real valued Gaussian function.

\noindent \textbf{Remark.} As a special limit case, we observe that in the classical limit, \ie $D_{qq}=D_{pq}=0$, the equilibrium state
simplifies to the well known stationary solution $w_\infty ^{cl}$ of the classical kinetic Fokker-Planck equation (\ref{fp}) with a harmonic
oscillator potential (\cf \cite{Ri}), \ie
\begin{equation}
\label{cstst}w^{cl} _{\infty }(x,\xi )= \frac{\omega _0\gamma}{(2\pi)^d D_{pp} }\ e^{-\frac{\gamma}{D_{pp}} 
\ (|\xi |^2+\omega
_0^2|x|^2)}.
\end{equation}
\vskip 0.2 cm
\noindent Note that, in contrast to the classical case, the quantum steady state is \emph{not} a function of the
classical energy 
\begin{equation}
H(x,\xi ):=\frac 12|\xi |^2+\frac12\omega _0^2|x|^2. 
\end{equation}
For the classical kinetic FP equation, the fact that $w_{\infty}^{cl} =w_{\infty}^{cl}(H)$, implies that the transport and the classical FP scattering 
operator vanish independently when applied to $w^{cl}_\infty $. 
In our quantum mechanical framework this is no longer true, since there the steady state $w _\infty $ results from a cancellation of the transport operator
and the scattering term $\mathcal L_{q}$.


\section{Long time behavior}

\noindent We are now in the position to describe the long time behavior of the linear QFP equation in the cases were both friction 
and an external potential of harmonic oscillator type are present, and in
cases where one is missing. We will start with the  latter situation.

\subsection{Dispersion estimates in the unconfined case}

\noindent We want to address the unconfined or dispersive cases, \ie either
$\gamma = 0$ or $\omega_0 = 0$ (or both).  By comparison with the
classical FP equation, we expect that the particles escape to infinity
and thus that the macroscopic density $n$ decays to $0$ as $t \to
\infty$. More precisely we have the following theorem.

\begin{theorem} Let be either $\gamma =0$ or $\omega _0 = 0$, or both, and
assume (\ref{li}), (\ref{hop}):
\begin{enumerate}
\item If $\ w_0\in L^1\cap L^2$, then the solution $w$ of the QFP equation satisfies
\begin{equation}
\label{wlp} 
{\| \ w(t) \ \|}_p \ \leq \ C_p \, R_w(t) ^{-\frac{d}{2q}} \ {\| \ w_0 \ \|}_1 \ , \qquad 1\leq p \leq \infty,
\end{equation}
where $C_p$ is a constant independent of $w_0$, $p^{-1}+q^{-1}=1$ and
\begin{equation}
\label{dw}
R_w (t) : = e^{-4 \gamma t} (4\lambda(t)\nu(t)-\mu^2(t)),
\end{equation}
which is a pointwise positive function with $R_w (t) \to \infty$ as $t \to \infty$.
\item Consequently we have for the corresponding density, that
\begin{equation}
\label{nlp} 
{\| \ n(t) \ \|}_p \ \leq \ C_p \, R_n(t)^{-\frac{d}{2q}} \ {\| \ w_0 \ \|}_1 \ , \qquad 1\leq p \leq \infty,
\end{equation}
where again the rate $R_n (t) \to \infty$ as $t \to \infty$. Explicitly 
\begin{equation}
R_n(t) := 2 \left(\lambda (t)\tilde \alpha^2(t) + \mu (t)\tilde \alpha(t) \tilde \beta(t) + \nu (t)\tilde \beta^2(t)\right),
\end{equation}
using 
\begin{equation*}
\tilde \alpha (t):=d^{-1}\diverg_{x_0} \left(X_t(x_0,\xi _0)\right), \quad \tilde \beta (t):=d^{-1}\diverg_{\xi _0} \left(X_t(x_0,\xi _0)\right).
\end{equation*}
\end{enumerate}
\end{theorem} 
\begin{proof}
The estimate in claim no. 1 is trivial in the case $p=\infty$ from the formula of the 
fundamental solution in (\ref{vartr}) and (\ref{ws}). The estimate 
for $p=1$ is due to the following property of the Green's function
\begin{equation*}
\int\int_{\mathbb R^d\times \mathbb R^d} G(t,x,\xi,x_0,\xi_0) dx_0 \, d\xi_0 \ = \ 1
\end{equation*}
and a trivial estimate over (\ref{ws}). The estimates for $1<p<\infty$ 
are then obtained by interpolation. The function $R_w (t)$ can be computed explicitly in each of the three cases and is given by:
\begin{enumerate} 
\item If $\ \omega _0>\gamma=0$, then with $\varphi=\varphi(t)=2\,t\,\omega_0$,
\begin{eqnarray*} \kern -15pt
R_w (t) &\!\!=\!\!&\frac
14\,\left(D^2_{qq}+\frac 1{\omega_0^4}\,D^2_{pp}\right)\,\left( \varphi^2
+2\cos
\varphi-2 \right) \\  && +\, {\frac 2{\omega_0^2}\,
D^2_{pq}}\,\left(\cos\varphi-1 \right)\,+\, \frac 1{2\,\omega_0^2}
\,D_{pp}\,D_{qq}\,\left(
\varphi^2-2\cos \varphi+2\right) . 
\end{eqnarray*} 
\item If $\gamma>\omega _0 =0$, then with $\chi=\chi(t)=e^{-2\,t\,\gamma}$,
\begin{eqnarray*} \kern -15pt
R_w (t) &\!\!=\!\!&
\frac
1{4\,\gamma^{\,4}}\,(D_{pp}^2\,+\,4\,\gamma\,D_{pp}\,D_{pq})\left(\,t\,\gamma\,(1-\chi^2)
-(1-\chi)^2\right)\\
 &&-\frac 1{\gamma^2}\,D^2_{pq}(1-\chi)^2\,+\,\frac t\gamma\,D_{qq}
\,(1-\chi^2)
\end{eqnarray*} 
\item If $\gamma=\omega _0 =0$, then 
\begin{equation*} 
R_w (t) ={  - \ 4\,{D^2_{pq}}\,t^2 + 4\,D_{pp}\,D_{qq}\,t^2 + \frac
1{3}\,{{D^2_{pp}}\,t^4}}\ . 
\end{equation*}
\end{enumerate} 
In all cases, $R_w$ diverges as $t \to \infty$ and thus claim no. 1 is proved.\newpar 
To prove claim no. 2 we first compute the integral of the fundamental solution w.r.t. $\xi $, which gives 
\begin{equation} 
\label{intxi} \int_{\mathbb R^d} G (t,x,\xi ,x_0,\xi _0) \,d\xi \ = R_n^{-d/2}(t) \ {\cal N} 
\left(\frac{x - X_t(x_0,\xi _0)}{\sqrt{R_n(t)}}\right), 
\end{equation} 
where $X_t$ is defined as in lemma 3.1 and 
\begin{equation*} {\cal N} (\sigma) := \ (2\pi)^{-d/2} \exp\left( -\frac{\sigma^2}2 \right) . 
\end{equation*} 
We can then deduce a formula for the evolution of the macroscopic density, using (\ref{vartr}), (\ref{ws}) 
and the above computation (\ref{intxi}), to obtain 
\begin{equation*} 
n(t,x) \ = \ R_n^{-d/2}(t) \int\int_{\mathbb R^d\times \mathbb R^d} {\cal N} 
\left( \frac{x - X_t(x_0,\xi _0))}{\sqrt{R_n(t)}} \right) \, w_0 (x_0,\xi_0) \, dx_0 \, d\xi_0 . 
\end{equation*} 
The dispersion estimates are then straightforward from this expression. 
On the other hand, the function $R_n (t)$ again can be computed explicitly: 
\begin{enumerate} 
\item If $\ \omega _0>\gamma=0$, then with $\varphi=\varphi(t)=2\,t\,\omega_0$,
\begin{equation*} 2\,\omega_0^3\,R_n (t)  \!=\!{  2\,\omega_0\, D_{pq}\left( 1 - 
{\cos \varphi} \right) + D_{pp}\left( \varphi - 
{\sin \varphi}  \right) +\omega_0^2\, D_{qq}\left( \varphi + {\sin \varphi} \right). }
\end{equation*} 
\item If $\gamma>\omega _0 =0$, then with $\chi=\chi(t)=e^{-2\,t\,\gamma}$,
\begin{equation*} 8 \gamma^3 R_n (t)  \!=\! 
{ D_{pp} \left(  4\chi  +4\gamma t-3 - \chi^2   \right)
+16 \gamma^3  t D_{qq} + 
8\gamma D_{pq} \left( \chi + 2\gamma t -1\right).}
\end{equation*}
\item If $\gamma=\omega _0 =0$, then 
\begin{equation*} 
R_n (t) \!=\! 2\,D_{qq}\,t + 2\,D_{pq}\,t^2 + \frac 23\,D_{pp}\,t^3. 
\end{equation*} 
\end{enumerate} 
In all three cases $R_n$ diverges as $t \to \infty$ and thus claim no. 2
is also proved. \end{proof} 
\noindent The behavior at $t=+\infty$ of the rate functions can be obtained directly from the explicit computations in the previous result, 
depending on the diffusion constants. In the following corollary we will consider only the physical most important case, namely $D_{pp}>0$. 
\begin{corollary} Assume $D_{pp}>0$ and (\ref{li}), (\ref{hop}), then we have, as $t\rightarrow +\infty$: 
\begin{enumerate} 
\item If $\ \omega _0>\gamma=0$, then $R_w(t)=O(t^2)$ and $R_n(t)=O(t)$.
\item If $\ \gamma>\omega _0 =0$, then $R_w(t)=O(t)$ and $R_n(t)=O(t)$.
\item If $\ \gamma=\omega _0 =0$, then $R_w(t)=O(t^4)$ and $R_n(t)=O(t^3)$. 
\end{enumerate} 
\end{corollary}
\vskip 0.5 cm 
\noindent \textbf{Remark.} The computation of $n$ can also be used to
measure the convergence rates towards the steady state  macroscopic
density $n_\infty$ in the confined case. This computation however is
quite involved and we  leave the details to the reader, since we will
take a more elegant approach in the next subsection.\newpar  Nevertheless
the advantage of such an explicit calculation would be that one can prove
exponential decay of the solution,  even for the classical FP equation 
(this can also be seen by comparison with spectral theoretical approaches
\cite{Ri}), whereas the entropy method used in section 4.2 only gives a
suboptimal rate in this case \cite{DeVi} .

\subsection{Exponential decay towards equilibrium}

We now assume the presence of friction and of the confining potential. 
Like in the classical case, we expect exponential decay of the solution
towards the equilibrium state, which is usually proved using spectral
theory. \newpar  However in this work we shall follow a different
approach,  the so-called \emph{entropy-entropy-dissipation method}
for classical FP type equations (see \cite{AMTU} and references
therein).  As we shall see it can be successfully applied in our quantum
mechanical context too (for an overview of the classical applications, see
\cite{MaVi}).

\noindent First let us rewrite the drift-vector $P$ given by (\ref{dp}) in the following way 
\begin{equation}
P(x,\xi )=D(\nabla A(x,\xi )+F(x,\xi )),
\end{equation} 
where the gradient is taken w.r.t. $(x,\xi )\in\mathbb R^{2d}$ and $A$ is defined as the normalized potential appearing in the expression of the equilibrium state, 
more precisly
\begin{equation}
\label{qstre}w _\infty (x,\xi )\equiv \exp\left(-A(x,\xi )\right).
\end{equation}
Consequently the QFP equation (\ref{qfph}) takes the standard form of a \emph{non-sym}\-\emph{metric drift-diffusion equation}
\begin{align}
\label{dde}&\partial _t w = \ \diverg\left[D( \nabla w +w (\nabla A + F))\right],
\quad t\in\mathbb R^+\\ 
\label{ic}&w (t=0,x,\xi )= \ w _0(x,\xi ).
\end{align}
The vector field $F$ is explicitly given by
\begin{align}
\label{fdef}F(x,\xi )& := D^{-1} P(x,\xi)-\nabla A(x,\xi ),
\end{align}
where $D^{-1}$ is the inverse of the diffusion matrix, $A$ is defined by (\ref{qstre}) and $P$ is the drift given in (\ref{dp}).

\noindent The rewritten QFP equation (for harmonic oscillator potentials) with equilibrium state given by (\ref{qstre}) 
can now be identified as a special case of FP type equations, see \cite{AMTU}. 
Therefore we introduce the concept of \emph{relative entropies} in the same sense as in the quoted work:
\begin{definition} Let $\mathcal A$ be either $\mathbb R$ or $\mathbb R^+$. Let $\varphi \in C(\bar {\mathcal A})\cap C^4(\mathcal A)$ satisfy
\begin{align}
\varphi (1)=0,\ \varphi ''\geq 0 \mbox{ with } \varphi ''\not=0 \mbox{ and } (\varphi ''')^2\leq \frac{1}{2}\varphi ''\varphi ^{IV}
\mbox{ on $\mathcal A$.}
\end{align}
Assume moreover that $f\in L^1(\mathbb R^n)$, $g\in L^1_+(\mathbb R^n)$
with
$\int_{\mathbb R^n} f \ dx=\int _{\mathbb R^n} g \ dx=1$  and $f/g\in
\bar{ \mathcal A}$
$g(dx)$ a.e. Then 
\begin{equation}
\label{rent}e_\varphi (f\mid g):=\int_{\mathbb R^n} \varphi \left(\frac{f}{g}\right)g(dx)
\end{equation}
is an admissible relative entropy of $f$ w.r.t. $g$ and generating function $\varphi $.
\end{definition}
\noindent Two relative entropies which are frequently used are the \emph{logarithmic relative entropy}, associated to the generating function
\begin{equation}
\label{egenl}\varphi_{1}(\alpha ):=\alpha \ln \alpha -\alpha+1,\quad \alpha\in\mathbb R^+.
\end{equation}
and the \emph{quadratic relative entropy} with generator
\begin{equation}
\label{egenq}\varphi_2(\alpha):=k (\alpha-1)^2,\quad \alpha\in\mathbb R,\ k>0.
\end{equation}
\noindent Indeed it is known, see \cite{AMTU}, that (up to a positive
multiplicative constant) for every admissible entropy generator 
$\varphi$, there exist constants $K_1$, $K_2 >0$ such that for all
$\alpha\in\mathbb R^+$, it holds that 
\begin{equation}
K_1 \varphi _1(\alpha )\leq \varphi (\alpha )\leq K_2 \varphi _2(\alpha ).
\end{equation} 

\noindent \textbf{Remark.} 
The above defined entropies should not be confused with the quantum mechanical 
\emph{von Neumann entropy} $S:=-\tr(R \ln R )$, where $R$ is the density operator of the particle ensemble \cite{Th} 
(also note that in contrast to the physical convention the minus sign in front of (\ref{rent}) is dropped).\newpar 
There is however a notion of entropy for quantum states, called \emph{Wehrl entropy} \cite{We}, which is closely related
to the logarithmic entropy defined above (using the pointwise nonnegative \emph{Husimi transform} of $w$, see, \eg, \cite{LiPa})
and which can be interpreted as a measure of \emph{coherence and localization of quantum states}. 
For details, see \cite{AnHa}, \cite{GnZy}, \cite{SlZy} and the references given
therein.

\noindent Since the solution of the QFP equation in general is not pointwise positive, it seems that we need to look for 
an admissible entropy on all of $\mathbb R$. This would imply, see \cite{AMTU}, that the the only admissible entropy 
useful for our purpose is the quadratic one (\ref{egenq}). 
We circumvent this shortcoming by decomposing the initial condition $w_0$ into its negative and positive parts. More precisely we write
\begin{equation}
w_0(x,\xi )=w_0^+(x,\xi )-w_0^-(x,\xi )\mbox{ a.e.}
\end{equation}
where now $w_0^+$, $w_0^-$ are both non-negative functions with mass $M_0^+$, $M_0^-$ respectively. Clearly condition 
(\ref{nc}) implies
\begin{equation}
1=M_0^+-M_0^-.
\end{equation} 
Having in mind that the QFP equation is linear, we denote by 
\begin{equation}
w^{\pm}(t,x,\xi ):=\int\int_{\mathbb R^d\times \mathbb R^d}w _0^{\pm}(x_0,\xi _0) \ G (t,x,\xi ,x_0,\xi _0) \ dx_0d\xi _0
\end{equation}
the mild solution of (\ref{dde}) corresponding to $w_0^{\pm}$. Corollary
2.1 implies that $w^{\pm}(t,\cdot )\geq 0$ a.e. and  we shall now apply
the entropy-entropy-dissipation method to each of the two functions
$w^+$, $w^-$. Note that the steady state associated to 
$w^{\pm}(t,\cdot )$ is given by $w_\infty ^{\pm}=M_0^{\pm} w_\infty$. 

\noindent In order to use the results established in \cite{AMTU}, we first need to check
that the following property is fulfilled.
\begin{lemma} Assume $\gamma >0$, $\omega _0>0$ as well as (\ref{li}), (\ref{hop}), $y:=(x,\xi )\in\mathbb R^{2d}$. Then 
it holds that
\begin{equation}
\label{cond}\diverg_{y}\left(DFw _\infty \right)=0\quad on \ \mathbb R^{2d}
\end{equation}
where $F$ is defined by (\ref{fdef}).
\end{lemma}
\begin{proof} 
To prove the claim, we note, using equation (\ref{qstre}), that condition
(\ref{cond}) is equivalent to
\begin{equation*}
\diverg_y\left(D \nabla w_\infty + P w_\infty \right)=0,
\end{equation*}
which is the stationary QFP equation for potentials of the form (\ref{hop}). Thus the claim is true by definition of $A,F$ and $w_\infty $.
\end{proof}
\noindent With these preparations we can now state the main result of this section.
\begin{theorem} Assume $\gamma >0$, $\omega _0>0$ and assume (\ref{li})-(\ref{hop}).
Let the initial data $w _0$ be such that 
$w_0(x,\xi )=w_0^+(x,\xi )-w_0^-(x,\xi )$ a.e., $w_0^{\pm}\in\L^1\cap L^2$ and assume 
\begin{equation}
\label{bient}e_\varphi (w_0^{\pm}\mid M^{\pm}w_\infty )<\infty .\tag{\textbf{A4}}
\end{equation}
Then there exists a $\kappa >0$ such that 
\begin{equation}
\label{exdee}e_\varphi (w^{\pm}(t,\cdot )\mid M_0^{\pm} w_\infty )\leq 
e^{-2 \kappa t} e_\varphi (w_0^{\pm}\mid M_0^{\pm}w_\infty ), \quad t>0.
\end{equation}
As a consequence, the solution of (\ref{qfph}) converges exponentially towards the equilibrium state. More precisely it holds that:
\begin{enumerate}
\item If $e_{\varphi_1} (w_0^{\pm}\mid w_\infty )<\infty $, with $\varphi _1$ defined in (\ref{egenl}), then
\begin{equation}
{\| \ w (t,\cdot )-w_\infty \| }_{1}\ \leq C e^{-\kappa t} ,\quad C\in
\mathbb R^+,\; t>0.
\end{equation}
\item More generally, if $e_{\varphi_2} (w_0^{\pm}\mid w_\infty )<\infty $, with $\varphi _2$ defined in (\ref{egenq}), then
\begin{equation}
{\| \ w (t,\cdot )-w_\infty \| }_{p}\ \leq C e^{-\kappa t} \quad C\in
\mathbb R^+,\; t>0,
\end{equation}
with $1\leq p\leq 2$.
\end{enumerate}
\end{theorem}
\begin{proof}
\noindent The proof of the first claim (\ref{exdee}) is a consequence of the following \emph{convex Sobolev inequality} \cite{AMTU} 
(in which we use the notation $d\rho_\infty=\rho_\infty\,dx$)
\begin{equation*} 
\label{coso}\int_{\mathbb R^{2d}} \varphi \left(\frac{\rho }{\rho _\infty }\right)d\rho _\infty\leq
\frac{1}{2\,\kappa }\int_{\mathbb R^{2d}} \varphi^{\prime \prime} \left(
\frac{\rho }{\rho _\infty }\right) \left | \ \nabla 
\left(\frac{\rho }{\rho _\infty }\right)\right|^2 d\rho _\infty, 
\end{equation*}
which holds for every admissible entropy generator $\varphi $ (\cf definition 4.1), 
every function $\rho \in L^1(\mathbb R^{2d})$, with $M=M_\infty $ and  
$\rho _\infty := \exp(-A(x,\xi ))\in L^1_+(\mathbb R^{2d})$, with uniformly convex potential 
$A$, \ie
\begin{equation}
\frac{\partial ^2A}{\partial (x,\xi )^2}\geq \kappa _1 \Id{2d},
\end{equation}
where $\kappa _1>0$ and $\Id{2d}$ denotes the identity matrix in $\mathbb R^{2d}$.
Using this inequality, we can now estimate (recall that $w^\pm
_\infty=M^\pm w_\infty $)
\begin{equation*}
\int_{\mathbb R^{2d}} \varphi \left(\frac{w^\pm }{w^\pm _\infty }\right)dw^\pm _\infty\leq
\frac{1}{2\delta \kappa _1}\int_{\mathbb R^{2d}} \varphi^{\prime \prime} \left(\frac{w^\pm }{w^\pm _\infty }\right)\nabla ^\top 
\left(\frac{w^\pm }{w^\pm _\infty }\right)D\nabla \left(\frac{w^\pm }{w^\pm _\infty }\right)dw^\pm _\infty ,
\end{equation*}
where $\delta >0$ is the smallest eigenvalue of $D$ given by
\begin{equation}
\delta :=\frac{1}{2}\left( D_{pp}+D_{qq}-\sqrt{(D_{pp}-D_{qq})^2+4D^2_{pq}}\right),
\end{equation}
since by definition of $\delta $, it holds that $\delta \Id{2d}\leq D$. Provided lemma 4.1, the results of \cite{AMTU} imply the exponential decay of 
the relative entropy with a rate $\kappa \equiv \delta \kappa _1$ and thus (\ref{exdee}) is proved.\newpar 
Using the well known \emph{Csisz\'ar-Kullback inequality} \cite{AMTU}, \cite{Cs}, \cite{Ku}, we further obtain
\begin{equation*}
{\| \ w^{\pm} (t,\cdot )-M_0^{\pm} w_\infty \| }_{1}\leq K e^{-\kappa t}, \quad K\in\mathbb R^+, t>0. 
\end{equation*} 
Since by (\ref{nc}) we have $1=M_0^+-M_0^-$, this implies 
\begin{align*}
{\| w(t,\cdot )-w_\infty \| }_1&={\| w^+(t,\cdot )-w^-(t,\cdot )-(M_0^+-M_0^-)w_\infty \| }_1 \\
&\leq {\| w^+(t,\cdot )-M_0^+w_\infty\|}_1+{\| w^-(t,\cdot )-M_0^-w_\infty \| }_1\\
& \leq Ce^{-\kappa t},\quad C\in\mathbb R^+.
\end{align*}
For entropy generators of the form (\ref{egenq}), we get from (\ref{exdee})
 \begin{equation*}
{\| \ w^{\pm} (t,\cdot )-M_0^{\pm} w_\infty \| }_{2,\delta}\leq 
e^{-\kappa t}{\| \ w^{\pm} _0-M_0^{\pm}w_\infty \| }_{2,\delta} \quad \forall \ t>0,
\end{equation*} 
where ${\| \cdot \| }_{2, \delta}$ is the $L^2$-norm with weight $\delta =1/w_\infty$, \ie 
\begin{equation*}
{\| \ f \ \|}^2 _{2, \delta}:=\int\int_{\mathbb R^d\times\mathbb R^d}|f (x,\xi ) |^2 w^{-1}_\infty ( dx, d\xi). 
\end{equation*}
As above, the use of the triangle inequality allows us to conclude that exponential decay (with rate $\kappa $) of $w$ holds in this weighted
$L^2$-norm. Since $1/w_\infty \geq 1$ pointwise, we have ${\| \ f \ \| }_2\leq {\| \ f \ \|} _{2, \delta}$ 
(here we have used the Lindblad condition (\ref{li}) to show that the constant in front of
$1/w_\infty$ is indeed greater than 1).\newpar This implies
exponential decay in the usual $L^2$-norm and thus, by interpolation, we
obtain the exponential convergence  of $w$ towards the steady state in all
$L^p$-norms with $1\leq p\leq 2$.
\end{proof} 
\noindent It should be noted, that the main drawback of the above theorem is the fact that it only holds for 
initial data with positive and negative part bounded in relative entropy (\ref{bient}). 
Furthermore one should have in mind that the above theorem fails, if
$\gamma =0$ or $\omega _0=0$, since then $A$ is no longer uniformly
convex.  However we have already seen that in these cases no finite-mass
steady state, different from zero, exists.

\noindent The precise value of the rate $\kappa $ can be obtained by the following
transformation. Define a new potential $\tilde A$ by
\begin{equation}
\tilde A(\tilde x,\tilde \xi ):=A\left(\sqrt{D}\ (x,\xi )^\top \right), 
\end{equation}
where $\sqrt{D}$ is the square root of $D$ in the sense of positive definite matrices. Then, it holds that 
\begin{equation}
\frac{\partial ^2\tilde A}{\partial (\tilde x, \tilde \xi )^2}\geq \kappa \I,
\end{equation}
which gives the optimal rate $\kappa $ as the smallest eigenvalue of the Hessian of $\tilde A$. For details, see \cite{AMTU} or \cite{Ri}.

\noindent \textbf{Remark.} In contrast to our result, the convergence to the equilibrium state for classical kinetic FP equations (\ref{fp}) is, 
by similar entropy methods, obtained with a suboptimal rate of order
$O(t^{-\infty} )$, see \cite{DeVi}.  This is due to the fact, that for
the classical FP equation, in addition to the thermal equilibrium state
(\ref{cstst}),  there exist other states with zero entropy dissipation. 
Indeed explicit calculations show that $\kappa$ depends on $\hbar$ and
converges to $0$ as $\hbar \rightarrow 0$.

\noindent We hope to extend the above results to more general, maybe nonlinear, potentials $V$ in forthcoming works. 
Although quite simple, the presented harmonic oscillator case is expected to be an essential prerequisite in this analysis. 

\vskip 0.5 cm

\noindent\textbf{Acknowledgment}

\noindent{\small  The first author thanks A. Arnold and L. Neumann for helpful discussions.\newpar 
This work has been supported by the FWF Wissenschaftskolleg "\emph{Differentialgleichungen}", 
the \emph{"Wittgenstein Award 2000"} of P.M. and the FWF-Project P$14876$, "\emph{Fokker-Planck und mittlere Feldgleichungen}". 
Additional financial support has been obtained from the EU-TMR project, "\emph{Asymptotic Methods in Kinetic Theory}", the Spanish 
DGES project PB98-1281 and the austrian-french program Amadeus $02580$ QF on "\emph{Partial differential equations for modelling
semi-conductors.}"}


\end{document}